\documentclass[12pt]{article}
\usepackage{graphicx,amsmath}
\usepackage{units}

\newlength{\dinwidth}
\newlength{\dinmargin}
\def\lapproxeq{\lower .7ex\hbox{$\;\stackrel{\textstyle                                                    
<}{\sim}\;$}}                                                    
\def\gapproxeq{\lower .7ex\hbox{$\;\stackrel{\textstyle                                                    
>}{\sim}\;$}}                                                    
\def\be{\begin{equation}}                                                    
\def\ee{\end{equation}}                                                    
\def\bea{\begin{eqnarray}}                                                    
\def\eea{\end{eqnarray}}                                                    
\def\GeV{\rm GeV}

\def\sh{\hat s}
\def\sh2{{\hat s}^2}
\def\mr{\langle r \rangle}
\begin{document}                                                    
\titlepage                                                    
\begin{flushright}                                                    
IPPP/11/05  \\
DCPT/11/10 \\                                                    
\today \\                                                    
\end{flushright}

\vspace*{0.5cm}
\begin{center}                                                    
{\Large \bf Pomeron universality from identical pion correlations  }\\ 

{\Large \bf at the LHC}

\vspace*{1cm}
                                                   
V.A. Schegelsky$^{a}$, A.D. Martin$^b$, M.G. Ryskin$^{a,b}$ and V.A. 
Khoze$^{a,b}$ \\                                                    
                                                   
\vspace*{0.5cm}                                                    
$^a$ Petersburg Nuclear Physics Institute, Gatchina, St.~Petersburg, 188300, Russia \\            
$^b$ Institute for Particle Physics Phenomenology, University of Durham, Durham, DH1 3LE \\

\vspace*{1cm}                                                    
                                                    
\begin{abstract}                                                    
Bose-Einstein correlations of identical pions produced in high-energy $pp$ collisions at the LHC allow a probe of the Pomeron exchange mechanism.  The size of the domain which emits the pions depends on the multiplicity of events, but not on the collider energy. This confirms the universal structure of Pomeron exchange. The data at relatively low multiplicities indicate that the size of the source created by one-Pomeron exchange is much less than the size of the proton.

\end{abstract}                                                        
\vspace*{0.5cm}                                                    
                                                    
\end{center}

\section{Introduction}

The  radius of the proton-proton interaction can be measured in two different ways:
as the $t$-slope of the elastic $pp$-scattering or via the width of the peak in identical particle correlations. At first sight, these two methods do not seem related to each other. In this paper, we use a Regge approach to trace, and to exploit, the connection.

The {\it universal} shrinkage of the diffraction cone in elastic $pp$-, $\pi p$- and $Kp$-scattering, driven by the Pomeron, was already discovered at fixed-target CERN-SPS energies, starting with the CERN-Gatchina experiment \cite{pnpi}. The  behaviour of elastic scattering at high collider energies is well described by 
 a parametrization in which the elastic slope logarithmically grows with energy

\be
B_{\rm el}(s) = B_{0} + 2\alpha^{\prime}_P\ln(s/s_{0}).
\ee

For multiparticle production in proton-proton collisions one might expect that the size of the emission source, $r$, in high-energy inelastic interactions also increases with the collision energy. Moreover, within Regge theory (Pomeron approach) we expect {\it universal} behaviour of this radius. To measure the value of $r$  one may  use the Bose-Einstein correlations (BEC) between two identical pions. 

The high statistics of the LHC allows the space-time structure of high-energy interactions to be studied in detail. The standard way to determine the size of the interaction region is to measure the width of the peak at low $Q$ in BEC \cite{hbt,gfg,kop,bec} of two identical pions emitted with four momenta $p_1$ and $p_2$, where  $Q\equiv \sqrt{-(p_1-p_2)^2}$. That is, the Lorentz-invariant quantity $Q$ gives a measure of the proximity of the two identical particles in phase space. The first LHC data showing the BEC enhancements at low $Q$ have been reported by CMS for $pp$ collisions \cite{CMS1,CMS2} and by ALICE for $pp$ and for Pb-Pb collisions \cite{alicepp,ALICE1, ALICE2}. 

The data are presented as the ratio of the $Q$ distribution for pairs of identical particles to a reference distribution for pairs of particles which are expected to have no BEC effect
\be
R(Q)~=~(dN/dQ)/(dN_{\rm ref}/dQ).
\ee
The peak in the $R(Q)$ distribution is then fitted to the exponential $\exp(-Qr)$ (in CMS case) or to the Gaussian $\exp(-Q^2r^2)$ (ALICE) parametric form, where the size of the pion-pair emission region is characterised by an effective radius $r$.

\section{Pomeron mechanism for multiparticle production}
Here, we discuss the BEC effect in terms of the microscopic interaction, rather than in terms of collective thermodynamic variables. For high-energy $pp$ interactions, the total and elastic cross sections are well described in terms of  Regge theory by an effective Pomeron pole -- an universal object with its own internal structure which does not depend on the energy, $\sqrt{s}$, of the colliding protons. Thus in the one-Pomeron exchange approach we do not expect the spectra and density of secondaries in the central region, $dN/d^3p$ to depend on $s$. However, this naive expectation contradicts the recent LHC data. In particular, the height of the central plateau in $dN/dy$ is observed to grow by more than 60$\%$ in going from a collider energy of 0.9 TeV to 7 TeV \cite{cms,atlas}.  Within Regge Field Theory this growth is explained by mutli-Pomeron exchange contributions. Besides one-Pomeron, there may be contributions from two or more Pomeron exchanges. The mean number of Pomeron exchanges increases with energy, since at a higher energy we have a larger total cross section and, thus, a larger probability of an additional rescattering, or additional multiple interactions, described by the exchange of additional Pomerons. This leads to the growth of the multiplicity\footnote{In Monte Carlo generators this effect is described by the multiple interaction option, when a few pairs of incoming partons interact simultaneously, each interaction corresponding to the exchange of one Pomeron.} $dN/d\eta$. 

\begin{figure} 
\begin{center}
\includegraphics[height=4cm]{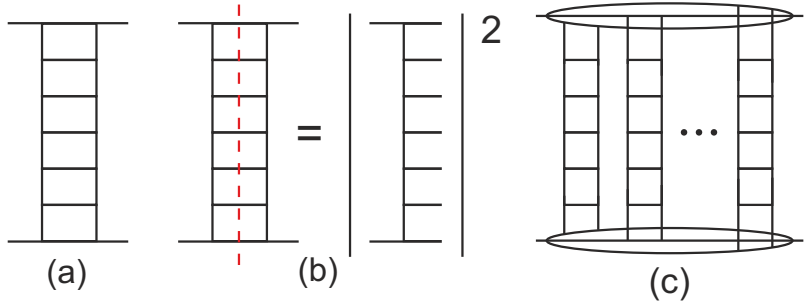}
\caption{\sf (a) The ladder diagram for one-Pomeron exchange; (b) cutting one-Pomeron exchange leads to the multiperipheral chain of final state particles; (c) a multi-Pomeron exchange diagram. }
\label{fig:abc}
\end{center}
\end{figure}

In the Regge approach each soft Pomeron can be regarded as a multiperipheral ladder, see Fig.~\ref{fig:abc}. The ladder shown in (a) corresponds to the elastic $pp$ scattering amplitude, while on cutting the Pomeron, as in (b), we obtain the cross section for multiparticle production.  Cutting $n$ Pomerons in a multi-Pomeron exchange diagram, (c), gives a final density of secondaries in the central region that is $n$ times larger than that for one-Pomeron exchange, see, for example, \cite{JPG} for more details.

\section{Radius of source depends on the number of Pomerons}
It is observed that the mean radius $\mr$, measured by identical particle correlations without multiplicity selection, grows with energy. This is naturally explained within the multi-Pomeron picture. Indeed, if we select events with low multiplicity then we deal with a process described by one-Pomeron exchange. In this case, the width of the correlation peak measures the size of the pion   source originating from one individual Pomeron. On the contrary, in high multiplicity events, combinatorial factors mean that the two identical pions are dominantly emitted from different Pomerons. Then the value of $\mr$ reflects the distance between the Pomerons.  Thus we expect $\mr$ will increase with the charged particle multiplicity, $N_{\rm ch}$, while the energy behaviour is explained by the growth of the number of exchanged Pomerons (and thus the number of secondaries) with $\sqrt{s}$. Indeed, the data \cite{CMS2,alicepp} show this kind of behaviour. We emphasize that the values of $\mr$ measured at quite different collider energies, 0.9 TeV to 7 TeV, are, to good accuracy, the same for the same value of $N_{\rm ch}$, that is, for the contribution from the same number of Pomeron exchanges. 

Such `scaling' behaviour (in which the value of $\mr$ does not depend on the proton energy, but on $N_{\rm ch}$ only) confirms the interpretation based on one, or more,  Pomerons exchanged. Thus the smallest radius  should be 
 considered as the radius of one individual Pomeron (or, to be more 
 precise, as the size of the source formed by a single Pomeron).
In other words, identical pion correlations at low  $N_{\rm ch}$ allow the measurement of the size of an individual Pomeron.

The same effect has been observed in heavy-ion collisions \cite{ALICE1,ALICE2}. The value of $\mr$ measured in events with a given particle density, $dN/d\eta$, turns out to be practically constant starting from very low fixed-target energies right up to the LHC energy $\sqrt{s_{NN}}$=2.76 TeV of the colliding constituent nucleons. In the case of heavy-ion collisions the final multiplicity depends on the centrality of the events. In other words, $dN/d\eta$ is proportional to the number of $NN$ collisions. For a very peripheral interaction, only one pair of nucleons collides, whereas for a central collision the discs of the heavy ions overlap completely, leading to the simultaneous interactions of many pairs of nucleons.

\section{Saturation of source radius at large $N_{\rm ch}$}

Let us return to $pp$ collisions, and recall the partonic structure of the proton. In a $pp$ collision the parton-parton interaction is analogous to a $NN$ collision in heavy-ion collisions. Thus we expect $\mr$ to saturate at the value of $N_{\rm ch}$ when the partonic discs of the colliding protons overlap, that is at a value of $\mr$ characteristic of the $pp$ interaction. Let us compare the value of $\mr \sim 2$ fm observed for large $N_{\rm ch}$ with that obtained from the $t$-slope of the elastic $pp$ cross section. The slope expected at 7 TeV is about $B_{\rm el} \sim 20 ~\GeV^{-2}$. Now the relation between the slope and the radius is\footnote{For the $t$-slope, $B_{\rm el}/2$, of the elastic amplitude, we deal with a two-dimensional transverse vector. Therefore we use $\langle r^2 \rangle/4$ instead of $\langle r^2 \rangle/6$.}
\be
B_{\rm el}/2~=~\langle r^2 \rangle/4~~~~{\rm gives}~~~\langle r^2 \rangle \sim 1.55~{\rm fm}^2.
\ee
However, the above relation is obtained with a Gaussian form $e^{-Q^2r^2}$, while the correlation is observed \cite{CMS1} to be better described by the exponential form $e^{-Qr}$.
%as shown in 
%(\ref{eq:form}).
 The relation between $\mr$ for the two forms is
\be
\mr_{\rm exp}~=~\sqrt{\pi}\mr_{\rm Gaussian}.
\ee
Thus we obtain $\mr_{\rm exp}=2.2$ fm close to the value found in the BEC data \cite{CMS2,ALICE2}.
That is, we have correspondence between the value of $\mr$ at large $N_{\rm ch}$  and the interaction radius determined via the elastic slope $B_{\rm el}$. Since $B_{\rm el}$ increases with energy, we therefore expect that the `saturated' BEC value of $\mr$ will also increase a little. 

\begin{figure} 
\begin{center}
\includegraphics[height=6cm]{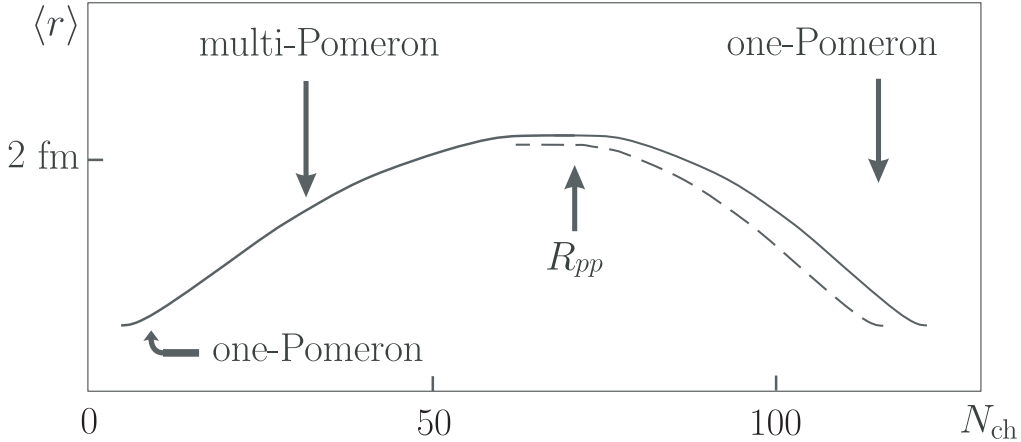}
\caption{\sf A sketch of the expected values of $\mr$ as a function of the multiplicity of charged particles, $N_{\rm ch}$. The size $\mr$ of the source of identical pions originating from different mechanisms is indicated. The continuous and dashed curves correspond to, say, $\sqrt{s}=$ 7 and 0.9 TeV respectively. At low $N_{\rm ch}$ we expect $\mr$ to be independent of the collider energy $\sqrt{s}$, while for the plateau we expect $\mr \sim R_{pp} \propto \sqrt{B_{\rm el}(s)}$ to increase very slowly with energy. Very large multiplicities are expected to arise from high-$E_T$ events originating from a single ladder. The scale for the $N_{\rm ch}$ axis depends on the experimental acceptance domain. Here, we indicate an approximate typical scale for $N_{\rm ch}$ corresponding to a relevant experimental $\eta$ interval, $|\eta|<2.5$, and $p_t$ cut. }
\label{fig:r}
\end{center}
\end{figure}

To conclude our discussion so far, we see that the BEC effects at the LHC offer the opportunity to confirm the universal small-size 
%(and almost point-like) 
 structure expected for the Pomeron\footnote{There were some old arguments in favour of a small transverse size of the Pomeron. First, the slope of the Pomeron trajectory, $\alpha'_P$, is much smaller than the slope $\alpha'_R$ observed for the trajectories of the secondary Reggeons, 
like $\rho,\ \omega,\ f_2,\ a_2$; recall that the value of $\alpha'$ is proportional to the square of the transverse size of the corresponding Reggeon.  Next,  the success of the additive quark model
 ($\sigma(\pi p)/\sigma(pp)\simeq 2/3$) is easy to explain when 
the radius of quark-quark interaction is much smaller than the hadron (pion, proton) radius. However now we have the first {\it direct} measurement of the size of the Pomeron. This result should not be affected by possible Pomeron-Pomeron interactions. The triple-Regge analysis indicates a very small radius of multi-Pomeron vertices, so the multi-Pomeron interactions do not noticeably change the radius of the pion source.} and to measure both the 
size of individual Pomeron (at comparatively small $N_{\rm ch}$) and the radius of the 
proton-proton interaction (at large $N_{\rm ch}$).
We sketch 
the expected behaviour of the average radius of the identical particle emission region, $\mr$ with $N_{\rm ch}$ in Fig.~\ref{fig:r}. At low $N_{\rm ch}$ we measure the size of the pion source created by one Pomeron, while at larger $N_{\rm ch}$ the result is mainly driven by the spatial separation between two Pomeron exchanges. An important result of the observation of BEC is that at low $N_{\rm ch}$ the size of the source does not depend on energy, within the experimental accuracy. This confirms the universal structure of the Pomeron pole.
At large $N_{\rm ch}$ we expect saturation of $\mr$ at a value corresponding to the radius, $R_{pp}$, of the $pp$ interaction. The corresponding plateau in the $N_{\rm ch}$-plot should therefore increase slowly with energy, as $\mr \propto \sqrt{B_{\rm el}(s)}$.

There is the possibility that very large multiplicity events will arise dominantly from high $E_T$ production, see point (iii) of Section \ref{sec:A5}. A high-$E_T$ process will mainly originate from one-ladder exchange, that is from one Pomeron\footnote{High $E_T$ cross sections decrease steeply with increasing energy, and it is better not to distribute the energy among a few Pomeron exchanges.}. Thus at very large $N_{\rm ch}$ we expect that the value of $\mr$ will decrease back towards the value due to the size of one Pomeron (or one jet); that is, close to the value of $\mr$ at low $N_{\rm ch}$.

\section{Complications    \label{sec:A5}}
So far our discussion has been idealistic. The actual situation is more complex. There are background sources which may complicate the interpretation of the BEC measurements. We divide these into three classes.

(i) Let us start with $\mr$ at small $N_{\rm ch}$. In the ideal case, this is just the radius of an individual Pomeron. Unfortunately here the radius is not well defined. In terms of the quark and gluon degrees of freedom the Pomeron, may be an almost point-like object with $\mr \sim 0$. However we need space to form a hadron (pion), which has its own size. In particular,  in $e^+e^-$ annihilations, where the interaction starts from the point-like production of a $q\bar{q}$ pair, an analogous radius is observed to be $\mr \sim 0.5-0.8 $ fm, depending on the form of the fit \cite{ee}. This is comparable to the radius, $R_{\pi}$, of the pion. Thus, even for a point-like Pomeron, we expect hadronization will give a non-zero $\mr \sim R_{\pi}$ for small $N_{\rm ch}$.

(ii) The next problem is that an appreciable fraction of pions is produced via resonance production and decay, rather than directly. For a long living resonance (with a small width $\Gamma$), the distance between the creation and the decay of the resonance, $\sim 1/\Gamma$, may be rather large leading to a larger $\mr$. The contribution, via resonances, is smaller for pions of larger transverse momentum $k_t$. Indeed, there is a tendency for $\mr$ to decrease with $k_t$ \cite{CMS2,ALICE2}.

(iii) On the other hand for large $k_t$, we have another `background' problem to consider.  Actually the value of $N_{\rm ch}$ measured at the LHC is not for the whole event. Instead, it is the number of tracks observed in the restricted region covered by the central tracking detectors. So by including the exchange of an additional Pomeron we only enlarge $N_{\rm ch}$ by some fixed amount. As we have mentioned in the previous section, high $E_T$-jet production is another possible origin of large $N_{\rm ch}$ events. Now jet multiplicity grows exponentially \cite{book}, $N_{\rm jet} \sim {\rm exp}(c\sqrt{{\rm ln}(E_T/q_0)})$, and the resulting particles are dominantly observed in the central detectors. Thus for very large $N_{\rm ch}$, there may be competition between contributions from high-$E_T$ jets and soft multi-Pomeron interactions. This background can be quantified by selecting events with one high $E_T$ particle and looking for BEC between two soft pions coming from the underlying event. In this way we may check whether the soft underlying event, which accompanies the hard process, has the same properties as normal soft multiparticle production.  If so, we may correct the data by eliminating the jet events. It looks, however, that 
%Of course, the observed saturation effect may well occur at not too lthe observed saturation effectarge values of $N_{\rm ch}$, with little jet contamination.
  the observed saturation effect already occurs at not too large values of $N_{\rm ch}$, where there is little jet contamination, see Fig.~\ref{fig:r}.

\section{Outlook}  
  Thanks to the large luminosity and enormous statistics available at the LHC, we now have the possibility to study the size of the interaction  volume, which produces the identical pion pair, under different kinematic conditions.
   In particular, we may measure:
  \begin{itemize}
\item  $\mr$ via the correlations of pions with, say,
  $\eta>1$ as a function of the multiplicity observed in the other hemisphere with $\eta<-1$. In this way we may check whether we deal with one source of secondaries or whether there are a few independent `fireballs'
  (that is, preconfinement clusters).
\item $\mr$ in the events with a high $E_T$ jet (or one high-$p_t$ hadron).
  Does high $E_T$ select the small-size component of the incoming proton?
  \item  $\mr$ in the events with a large rapidity gap. A small gap survival probability $S^2$ pushes the amplitude to the periphery,
  where there is a smaller probability of rescattering (and of an extra Pomeron exchange); for a recent review see \cite{ep}.
  \item $\mr$ as a function of the rapidity and/or the transverse momentum of the pion pair.
  \item $\mr$ in the correlations of kaons ($K^0_S$); and to compare with radius of the pion source.
  \end{itemize}

\section*{Acknowledgements}
MGR would like to thank the IPPP at the University of Durham for hospitality; and the Federal Program of the Russian State RSGSS-65751.2010.2  for support of grant RFBR
11-02-00120-a.


\thebibliography{}
\bibitem{pnpi} J.P. Burq et al., Phys. Lett {\bf B109}, 124 (1982).

\bibitem{hbt} R. Hanbury-Brown and R.W. Twiss, Phil. Mag. {\bf 45}, 663 (1954); Proc. Roy. Soc. {\bf 242A}, 300 (1957); {\it ibid} {\bf 243A}, 291 (1957).
\bibitem{gfg} G. Goldhaber, W.B. Fowler, S. Goldhaber et al.,
 Phys. Rev. Lett. {\bf 3}, 181 (1959).
\bibitem{kop}  G.I. Kopylov and M.I. Podgoretskii, Sov. J. Nucl. Phys. {\bf 15}, 219 (1972); {\bf 18}, 336 (1973).
\bibitem{bec} G. Alexander, Rep. Prog. Phys. {\bf 66}, 481 (2003).

\bibitem{CMS1} CMS Collaboration: V. Khachatryan et al., Phys. Rev. Lett. {\bf 105}, 032001 (2010).

\bibitem{CMS2} CMS Collaboration: V. Khachatryan et al., arXiv:1101.3518.
\bibitem{alicepp} ALICE Collaboration: K. Aamodt et al., arXiv:1101.3665.
\bibitem{ALICE1} ALICE Collaboration: K. Aamodt et al., arXiv:1007.0516. 
%Phys. Rev. {\bf D82}, 052001 (2010).

\bibitem{ALICE2} ALICE Collaboration: K. Aamodt et al., arXiv:1012.4035.
\bibitem{cms} CMS Collaboration:  V. Khachatryan et al., Phys. Rev. Lett. {\bf 105}, 022002 (2010). 
\bibitem{atlas} ATLAS Collaboration: Georges Aad et al., arXiv:1012.0791. 
\bibitem{JPG} M.G. Ryskin, A.D. Martin, V.A. Khoze and A.G. Shuvaev, J. Phys. G. {\bf 36}, 093001 (2009).

\bibitem{ee} see, for example, P. Abreu et al., Phys. Lett. {\bf B471}, 460 (2000).
\bibitem{book} Y.L.~Dokshitzer, V.A.~Khoze, A.H.~Mueller and S.I.~Troian,
`Basics Of Perturbative QCD', {\it  Gif-sur-Yvette, France: Ed. Frontieres (1991) 274 p.}
\bibitem{ep} A.D.~Martin, M.G.~Ryskin and V.A.~Khoze,
  %``Forward Physics at the LHC,''
  Acta Phys.\ Polon.\   {\bf B40}, 1841 (2009)
  [arXiv:0903.2980].
\end{document}